\pdfoutput=1
\documentclass[11pt]{article}
\usepackage{graphicx}
\usepackage{latexsym}
\usepackage{mathrsfs}
\usepackage[overload]{textcase}

\font\grande=cmr9.5 scaled \magstep4
\font\medio=cmr9.5 scaled \magstep2
\outer\def\beginsection#1\par{\medbreak\bigskip
      \message{#1}\leftline{\bf#1}\nobreak\medskip
\vskip-\parskip
      \noindent}

\begin{document}

\bibliographystyle{unsrt}

\titlepage

\vspace{1cm}
\begin{center}
{\grande Effective field theories and inflationary magnetogenesis}\\
\vspace{1.5 cm}
Massimo Giovannini \footnote{e-mail address: massimo.giovannini@cern.ch}\\
\vspace{1cm}
{{\sl Department of Physics, CERN, 1211 Geneva 23, Switzerland }}\\
\vspace{0.5cm}
{{\sl INFN, Section of Milan-Bicocca, 20126 Milan, Italy}}
\vspace*{1cm}
\end{center}
\vskip 0.3cm
\centerline{\medio  Abstract}
\vskip 0.5cm
The effective approach is applied to the analysis of inflationary 
magnetogenesis. Rather than assuming a particular underlying description, all the generally 
covariant terms potentially appearing with four space-time derivatives in the effective action have 
been included and weighted by inflaton-dependent couplings.  The higher derivatives are 
suppressed by the negative powers of a typical mass scale whose specific values ultimately depend on the tensor to scalar ratio. During a quasi-de Sitter stage the corresponding corrections always lead to an asymmetry between the hypermagnetic and the hyperelectric susceptibilities. After presenting a general method for the estimate of the gauge power spectra, the obtained results are illustrated for generic models and also in the case of some non-generic scenarios where either the inflaton has some extra symmetry or the higher-order terms are potentially dominant.
\noindent
\vspace{5mm}
\vfill
\newpage

The dynamical evolution of a large class of inflationary models is conventionally 
described in terms of a scalar-tensor theory of gravity whose effective Lagrangian density is characterized by a single 
inflaton field $\varphi$ 
\begin{equation}
{\mathcal L}_{inf} = \sqrt{- G} \biggl[ - \frac{ \overline{M}_{P}^2\,\, R}{ 2} + 
\frac{1}{2} G^{\alpha\beta} \partial_{\alpha} \varphi \partial_{\beta} \varphi - V(\varphi)\biggr],
\label{ONE}
\end{equation}
where $\overline{M}_{P} = M_{P}/\sqrt{8 \pi}$ is the reduced Planck mass
while $V(\varphi)$ denotes the inflaton potential. Equation  (\ref{ONE}) can be regarded as  the first term of a 
generic effective field theory where the higher derivatives are suppressed by the negative powers of a large mass $M$ associated with the fundamental theory that underlies the effective description. 
For practical reasons it will be useful to deal with an appropriate dimensionless scalar $\phi = \varphi/M$. Following the lucid discussion of Ref. \cite{ONE} (see also \cite{ONEa}) the leading correction to Eq. (\ref{ONE}) consists of all possible terms containing four derivatives and it can be parametrized in the following manner: 
\begin{eqnarray}
&& \Delta\,{\mathcal L}_{inf} = \sqrt{-G} \biggl[ c_{1}(\phi) \bigl(G^{\alpha\beta} \partial_{\alpha} \phi \partial_{\beta} \phi\bigr)^2 
+ c_{2}(\phi) G^{\mu\nu}\, \partial_{\mu} \phi \, \partial_{\nu} \phi\, \Box \phi + 
c_{3}(\phi) \bigl( \Box \phi \bigr)^2 
\nonumber\\
&& + c_{4}(\phi) \, R^{\mu\nu} \, \partial_{\mu} \phi \,\partial_{\nu} \phi
+ c_{5}(\phi) \, R\, G^{\mu\nu} \,\partial_{\mu} \phi \,\partial_{\nu} \phi
+ c_{6}(\phi) R \, \Box \phi + c_{7}(\phi) R^2 + c_{8}(\phi) \, R_{\mu\nu} \,R^{\mu\nu} 
\nonumber\\
&& + c_{9}(\phi) R_{\mu\alpha\nu\beta} \, R^{\mu\alpha\nu\beta} + c_{10}(\phi) C_{\mu\alpha\nu\beta} \, C^{\mu\alpha\nu\beta}
+ c_{11}(\phi)  R_{\mu\alpha\nu\beta} \, \widetilde{\,R\,}^{\mu\alpha\nu\beta} + c_{12}(\phi) C_{\mu\alpha\nu\beta} \, \widetilde{\,C\,}^{\mu\alpha\nu\beta}\biggr],
\label{THREE}
\end{eqnarray}
where $R_{\mu\alpha\nu\beta}$ and $C_{\mu\alpha\nu\beta}$ denote the Riemann and  Weyl tensors while 
$ \widetilde{\,R\,}^{\mu\alpha\nu\beta}$ and $\widetilde{\,C\,}^{\mu\alpha\nu\beta}$ are the corresponding duals. From Eq. (\ref{THREE}) various interesting conclusions can be drawn. For instance the leading correction to the two-point function of the scalar mode of the geometry comes from the terms containing four-derivatives of the inflaton field while in the case of the 
tensor modes the leading corrections stem from  $C_{\mu\alpha\nu\beta} \,\widetilde{C}^{\mu\alpha\nu\beta}$ and $R_{\mu\alpha\nu\beta} \,\widetilde{R}^{\mu\alpha\nu\beta}$ which are typical of Weyl and Riemann gravity \cite{TWO,THREE}. 
Incidentally both terms break parity and are therefore capable of polarizing the stochastic backgrounds of the 
relic gravitons \cite{FOUR} by ultimately affecting the dispersion relations of the two circular polarizations.

The same logic shall now be extended to the description of magnetogenesis scenarios based on the evolution 
of the gauge coupling so that the Lagrangian density (\ref{ONE}) 
will now be complemented by the contribution of the hypercharge fields:
\begin{equation}
{\mathcal L}_{gauge} =- \sqrt{- G} \biggl[ \frac{\lambda(\phi)}{16\pi} Y_{\alpha\beta} \,\, Y^{\alpha\beta} +\frac{\overline{\lambda}(\phi)}{16\pi} Y_{\alpha\beta} \, \widetilde{\,Y\,}^{\alpha\beta} \biggr],\qquad\qquad \widetilde{\, Y\,}^{\alpha\beta} = E^{\alpha\beta\mu\nu} Y_{\mu\nu}/2,
\label{FOUR}
\end{equation}
where  $E^{\alpha\beta\mu\nu} = \epsilon^{\alpha\beta\mu\nu}/\sqrt{-G}$ and
$\epsilon^{\alpha\beta\mu\nu}$ is the four-dimensional Levi-Civita symbol;
within the notations of Eq. (\ref{FOUR}) the gauge coupling is  $g = \sqrt{4\pi/\lambda}$.  For the Lagrangian density  
 Eq. (\ref{FOUR}) the analog of  $\Delta\, {\mathcal L}_{inf}$ will have to include the collection of all possible terms containing four derivatives and combining the inflaton, the gauge fields and the metric tensor: 
\begin{eqnarray}
&& \Delta {\mathcal L}_{gauge} = \frac{\sqrt{-G}}{16 \, \pi\, M^2} \biggl[ \lambda_{1}(\phi) \, R\, Y_{\alpha\beta}\, Y^{\alpha\beta} + \lambda_{2}(\phi) \, R_{\mu}^{\,\,\,\,\,\nu} \, Y^{\mu\alpha}\, Y_{\alpha\nu}  + \lambda_{3}(\phi) \, R_{\mu\alpha\nu\beta} \, 
Y^{\mu\alpha} \, Y^{\nu\beta} 
\nonumber\\
&& + \lambda_{4}(\phi) \, C_{\mu\alpha\nu\beta} \, Y^{\mu\alpha} \, Y^{\nu\beta} + \lambda_{5}(\phi) \Box \phi \,\, Y_{\alpha\beta} \, Y^{\alpha\beta}
+ \lambda_{6}(\phi) \partial_{\mu}\phi \partial^{\nu}\phi 
Y^{\mu\alpha} \, Y_{\nu\alpha} + \lambda_{7}(\phi) \nabla_{\mu}\nabla^{\nu} \phi\,  Y_{\nu\alpha} Y^{\mu\alpha}
\nonumber\\
&& + \overline{\lambda}_{8}(\phi) \, R\, Y_{\alpha\beta}\, \widetilde{\, Y\,}^{\alpha\beta} + \overline{\lambda}_{9}(\phi) \, R_{\mu}^{\,\,\,\,\,\nu} \, Y_{\alpha\nu}  \widetilde{\, Y\,}^{\mu\alpha}\, + \overline{\lambda}_{10}(\phi) \, R_{\mu\alpha\nu\beta} \, 
Y^{\mu\alpha} \, \widetilde{\, Y\,}^{\nu\beta} + \overline{\lambda}_{11}(\phi) \, C_{\mu\alpha\nu\beta} \, Y^{\mu\alpha} \, \widetilde{\,Y\,}^{\nu\beta} 
\nonumber\\
&& + \overline{\lambda}_{12}(\phi) \Box \phi \,\, Y_{\alpha\beta} \, \widetilde{\,Y\,}^{\alpha\beta}
+ \overline{\lambda}_{13}(\phi) \partial_{\mu}\phi \partial^{\nu}\phi 
\widetilde{\,Y\,}^{\mu\alpha} \, Y_{\nu\alpha} + \overline{\lambda}_{14}(\phi) \nabla_{\mu}\nabla^{\nu} \phi\,  Y_{\nu\alpha} \, \widetilde{\,Y\,}^{\mu\alpha}
 \biggr].
\label{FIVE}
\end{eqnarray}
Equation (\ref{FIVE}) contains $14$ distinct terms; $7$ of them do not break parity and are weighted by the couplings $\lambda_{i}(\phi)$ (with $i =1,\,\, ...\,,\,\,7$). The remaining $7$ contributions are weighted by the prefactors $\overline{\,\lambda\,}_{j}(\phi)$ (with $j =8,\,\, ...\,,\,\,14$) and contain parity-breaking terms. The contributions containing the dual Riemann or Weyl tensors 
(e.g. $\widetilde{R}_{\mu\alpha\nu\beta} Y^{\mu\alpha} Y^{\nu\beta}$ and $\widetilde{C}_{\mu\alpha\nu\beta} Y^{\mu\alpha} Y^{\nu\beta}$) are fully 
equivalent to the ones already present in Eq. (\ref{FIVE}) by recalling the explicit definition\footnote{The same kind of comment 
holds for terms containing a pair of dual fields of different nature (e.g. $\widetilde{R}_{\mu\alpha\nu\beta} Y^{\mu\alpha} \widetilde{Y}^{\nu\beta}$); in these cases the resulting expression will ultimately 
contain two four-dimensional Levi-Civita symbols whose contraction leads to a string of contributions that are already 
included in Eq. (\ref{FIVE}).} of $\widetilde{R}_{\mu\alpha\nu\beta}$ and $\widetilde{C}_{\mu\alpha\nu\beta}$.  Various particular case implicitly contained in Eq. (\ref{FIVE}) have been separately discussed in specific physical contexts but they 
have never been concurrently studied together with the inflaton coupling. For instance when the $\phi$-dependent couplings disappear (i.e. $\lambda_{i}(\phi) \to 1$) the first three terms have been analyzed by Drummond 
and Hathrell \cite{FIVE} in the curved version of quantum electrodynamics. Always in the absence of scalar couplings the considerations of Ref. \cite{FIVE} have been applied to the analysis of large-scale magnetism long ago mostly with negative conclusions. More recently the same terms (in a similar approximation) have been considered 
in Ref. \cite{EIGHTa} for the analysis of photon propagation in curved space-times.
Even more recently the Riemann coupling associated with $\overline{\lambda}_{10}(\phi)$ has been proposed
 in Ref. \cite{FOUR}; this term may ultimately polarize the relic graviton background.
The contributions containing the gradients of the inflaton (i.e. $\lambda_{5}(\phi)$, $\lambda_{6}(\phi)$, $\lambda_{7}(\phi)$ and their corresponding duals) arise in the relativistic theory of Van der Waals (or Casimir-Polder) interactions in flat \cite{SEVEN,EIGHT}
and curved \cite{TWELVE} backgrounds. 
It is finally appropriate to stress that we shall be interested in the situation where the gauge 
fields are amplified from their quantum fluctuations so that
the gauge fields vanish on the background and Eq. (\ref{FIVE}) does not include terms like 
$(Y_{\mu\nu}\, Y^{\mu\nu})^2$ typically appearing in the Euler-Heisenberg Lagrangian. When a classical gauge field background is  present these terms should be however included and may play a relevant role as argued\footnote{While the inclusion of a gauge background is clearly contrary to the logic of magnetogenesis (where the gauge fields should be dynamically generated)  it is 
interesting to remark the the effects on the effective gauge couplings are somehow similar to the ones 
produced by the inflaton background and by the geometry.} in Ref. \cite{TWELVEa}. 

The full Lagrangian density ${\mathcal L}_{gauge} + \Delta {\mathcal L}_{gauge}$ encompassing Eqs. (\ref{FOUR}) 
and (\ref{FIVE}) does not necessarily imply that the electric and magnetic susceptibilities must coincide. 
Let us in fact  consider, for the sake of concreteness, a conformally 
flat background geometry  $\overline{g}_{\mu\nu} = a^2(\tau) \, \, \eta_{\mu\nu}$
where $\eta_{\mu\nu}$ is the Minkowski metric and $a(\tau)$ is the scale factor 
written in terms of the conformal time coordinate $\tau$. In this case the full gauge action is:
\begin{equation}
S_{gauge} = \int d^{3} x \int d\tau \biggl({\mathcal L}_{gauge} + \Delta {\mathcal L}_{gauge}\biggr) 
= \frac{1}{2}\int d^{3} x \int d\tau \biggl(\chi_{E}^2 \, E^2 - \chi_{B}^2 B^2 + \overline{\chi}^2 \vec{E}\cdot\vec{B} \biggr),
\label{SIX}
\end{equation}
where $\vec{E}$ and $\vec{B}$ denote the comoving fields that are related to their physical counterparts as\footnote{In terms of the physical fields we obviously have $Y_{0 i} =  a^2 E_{i}^{(phys)}$ and  $Y^{i\, j} = - \epsilon^{i j k} B_{k}^{(phys)}/a^2$.}
$\vec{B} = a^2\,\, \chi_{B} \,\, \vec{B}^{(phys)}$ and as $\vec{E} = a^2 \,\,\chi_{E} \,\, \vec{E}^{(phys)}$. The hyperelectric and the hypermagnetic susceptibilities $\chi_{E}^2$, $\chi_{B}^2$ and $\overline{\chi}^2$ are instead defined as:
\begin{eqnarray}
\chi_{E}^2  &=& \frac{\lambda}{4\pi} \biggl\{ 1 + \frac{6 \, ({\mathcal H}^2 + {\mathcal H}^{\prime})}{M^2\, \, a^2} 
\biggl(\frac{\lambda_{1}}{\lambda}\biggr) + \frac{({\mathcal H}^2 + 2\,{\mathcal H}^{\prime})}{M^2\, \, a^2} 
\biggl(\frac{\lambda_{2}}{\lambda}\biggr) + \frac{2\,{\mathcal H}^{\prime}}{M^2\, \, a^2} 
\biggl(\frac{\lambda_{3}}{\lambda}\biggr) 
\nonumber\\
&-& \frac{\,(\phi^{\prime\prime} + 2 {\mathcal H} \phi^{\prime})}{M^2\, \, a^2} \biggl[ \biggl(\frac{\lambda_{5}}{\lambda}\biggr) 
+ \frac{1}{2} \biggl(\frac{\lambda_{6}}{\lambda}\biggr)\biggr] 
- \frac{\phi^{\prime\, 2}\, }{2\, M^2}\biggl(\frac{\lambda_{7}}{\lambda}\biggr)\biggr\},
\label{EIGHT}\\
\chi_{B}^2  &=& \frac{\lambda}{4\pi} \biggl[ 1 + \frac{6 \, ({\mathcal H}^2 + {\mathcal H}^{\prime})}{M^2\, \, a^2} 
\biggl(\frac{\lambda_{1}}{\lambda}\biggr) + \frac{({\mathcal H}^{\prime} + 2\,{\mathcal H}^{2})}{M^2\, \, a^2} 
\biggl(\frac{\lambda_{2}}{\lambda}\biggr) + \frac{2 {\mathcal H}^{2}}{M^2\, \, a^2} 
\biggl(\frac{\lambda_{3}}{\lambda}\biggr) 
\nonumber\\
&-& \frac{(\phi^{\prime\prime} + 2 {\mathcal H} \phi^{\prime})}{M^2\, \, a^2} \biggl(\frac{\lambda_{5}}{\lambda}\biggr) 
+ \frac{2\, {\mathcal H}\phi^{\prime}}{2\, M^2 a^2}\biggl(\frac{\lambda_{6}}{\lambda}\biggr)\biggr],
\label{NINE}\\
\overline{\chi}^2 &=& \frac{\overline{\lambda}}{4 \pi} \biggl\{ 1 + \frac{6 \,  ({\mathcal H}^2 + {\mathcal H}^{\prime})}{M^2\, \, a^2}
\biggl[ 6 \biggl(\frac{\overline{\lambda}_{8}}{\overline{\lambda}}\biggr) + \frac{3}{2} \biggl(\frac{\overline{\lambda}_{9}}{\overline{\lambda}}\biggr) + \biggl(\frac{\overline{\lambda}_{10}}{\overline{\lambda}}\biggr)\biggr]
\nonumber\\
&-& 2  \frac{(\phi^{\prime\prime} + 2 {\mathcal H} \phi^{\prime})}{M^2\, \, a^2} \biggl(\frac{\overline{\lambda}_{12}}{\overline{\lambda}}\biggr)
- \frac{(\phi^{\prime\prime} + 4 {\mathcal H} \phi^{\prime})}{M^2\, \, a^2} \biggl(\frac{\overline{\lambda}_{13}}{\overline{\lambda}}\biggr)
-  \frac{\,\phi^{\prime\, 2}\,}{4\, M^2 a^2}\biggl(\frac{\overline{\lambda}_{14}}{\overline{\lambda}}\biggr)\biggr\},
\label{TEN}
\end{eqnarray}
where the prime denotes a derivation with respect to the conformal time coordinate $\tau$, while, as usual, ${\mathcal H} = a H = a^{\prime}/a$. From Eq. (\ref{SIX}) the evolution equations for the hyperelectric and for the hypermagnetic fields are:
\begin{eqnarray}
&& \vec{\nabla}\times \bigl( \chi_{B} \, \vec{B} \bigr) - \partial_{\tau} \bigl(\chi_{E} \, \vec{E}\bigr) +
\biggl(\frac{\partial_{\tau}\overline{\chi}^2}{\chi_{B}} \biggr)\vec{B}  =0,\qquad \vec{\nabla} \cdot \bigl( \chi_{E} \, \vec{E} \bigr) +\vec{\nabla} \cdot \biggl[\biggl(\frac{\overline{\chi}^2}{\chi_{B}}\biggr) \vec{B}\biggr] =0,
\label{TENa}\\
&& \partial_{\tau} \biggl(\frac{\vec{B}}{\chi_{B}}\biggr) + \vec{\nabla} \times \biggl(\frac{\vec{E}}{\chi_{E}}\biggr) =0,\qquad \vec{\nabla}\cdot\biggl(\frac{\vec{B}}{\chi_{B}}\biggr) =0.
\label{TENb}
\end{eqnarray}
It is relevant to mention that when $\overline{\chi} \to  0$ (i.e. in the absence of parity-breaking terms) Eqs. (\ref{TENa}) and (\ref{TENb}) are invariant for a generalised duality symmetry:  when the susceptibilities are exchanged (i.e. $\chi_{B}\to 1/\chi_{E}$) the underlying equations are invariant provided  $\vec{B} \to \vec{E}$ and $\vec{E} \to - \, \vec{B}$.
In the limit $\chi_{E} \to \chi_{B}$ this is exactly the standard duality symmetry 
\cite{NINE,TEN} here analyzed in a conformally flat background.

While Eqs. (\ref{EIGHT}), (\ref{NINE}) and (\ref{TEN}) only assume a conformally flat background geometry, in view of the inflationary applications it is desirable to rephrase Eqs. (\ref{EIGHT}), (\ref{NINE}) and (\ref{TEN}) by introducing the slow-roll parameters  $\epsilon = - \dot{H}/H^2$ and $ \eta = \ddot{\phi}/(H\, \dot{\phi})$ (see for instance \cite{THIRTEENa,THIRTEEN}) and by also rescaling the gauge couplings; the result of this twofold manipulation is the following:
 \begin{eqnarray}
 && \chi_{E}^2  = \frac{\lambda}{4 \pi} \biggl( 1 + \frac{H^2 }{M^2} d_{E}^{(1)} - \epsilon \frac{H^2 }{M^2} d_{E}^{(2)} - \epsilon \frac{H^2}{M^2} d_{E}^{(3)} +  \sqrt{\epsilon} \frac{H^2 M_{P}}{M^3} d_{E}^{(4)} + \sqrt{\epsilon}\,\, \eta \, \frac{H^2 M_{P}}{M^3} \,d_{E}^{(5)}\biggr),
\label{ELEVEN}\\
&& \chi_{B}^2 = \frac{\lambda}{4 \pi} \biggl( 1 + \frac{H^2 }{M^2} d_{B}^{(1)} - \epsilon \frac{H^2}{M^2} d_{B}^{(2)} - \sqrt{\epsilon}\, \frac{H^2 M_{P}}{M^3} d_{B}^{(3)} +\sqrt{\epsilon} \,\, \eta\, \frac{H^2 M_{P}}{M^3} \,\, d_{B}^{(4)}\biggr),
\label{TWELVE}\\
&& \overline{\chi}^2 = \frac{\overline{\lambda}}{4\pi}\biggl( 1 + \frac{H^2}{M^2}  \overline{d}^{(1)} - \epsilon \frac{H^2 }{M^2} \overline{d}^{(2)} - \epsilon \frac{H^2 M_{P}^{2}}{M^{4}}  \overline{d}^{(3)}
+ \sqrt{\epsilon}\, \frac{H^2 M_{P}\,}{M^{3}}\, \overline{d}^{(4)} +  \sqrt{\epsilon} \,\,\eta \frac{M_{P} H^2 }{M^{3}}\overline{d}^{(5)}  \biggr).
\label{THIRTEEN}
\end{eqnarray}
Equations (\ref{THREE}) and (\ref{FIVE}) have in fact the same content since they represent the lowest terms of an expansion in inverse powers of $M$.
In Tab. \ref{TAB1} the explicit expressions of the $\phi$-dependent couplings  appearing in 
Eqs.(\ref{ELEVEN}), (\ref{TWELVE}) and (\ref{THIRTEEN}) have been collected by directly 
employing the Planck mass $M_{P}$ and not its reduced counterpart. 
\begin{table}
\begin{center}
\caption{The explicit definitions of the $\phi$-dependent couplings appearing $\chi_{E}$, $\chi_{B}$ and $\overline{\chi}$.}
\vskip 0.5truecm
\begin{tabular}{| c | | | c |}
\hline
$ d_{E}^{(1)} = 12 (\lambda_{1}/\lambda) + 3 (\lambda_{2}/\lambda) + 2 (\lambda_{3}/\lambda)$ & $ d_{E}^{(2)} = 6 (\lambda_{1}/\lambda) + 2 (\lambda_{2}/\lambda) + 2 (\lambda_{3}/\lambda)\qquad$ 
\\ \hline
$d_{E}^{(3)} =(\lambda_{7}/\lambda)/8 \pi\qquad\qquad\qquad\qquad\quad$  & $ d_{E}^{(4)} = [(\lambda_{5}/\lambda) + (\lambda_{6}/\lambda)/2]/\sqrt{4\pi} = 3\,d_{E}^{(5)}$
  \\ \hline 
$ d_{B}^{(1)} = 12 (\lambda_{1}/\lambda) + 3 (\lambda_{2}/\lambda) + 2 (\lambda_{3}/\lambda)\quad$ & $ d_{B}^{(2)} = 6 (\lambda_{1}/\lambda) + (\lambda_{2}/\lambda) \qquad\qquad\qquad\quad$ 
\\ \hline
$ d_{B}^{(3)} = [ 3(\lambda_{5}/\lambda)- (\lambda_{6}/\lambda)]/\sqrt{4\pi} \qquad\qquad$ & $d_{B}^{(4)} =(\lambda_{5}/\lambda)/\sqrt{4 \pi}\qquad\qquad\qquad\qquad\quad$
    \\ \hline
$ \qquad\overline{d}^{(1)} = 12 (\overline{\lambda}_{8}/\overline{\lambda}) + 3(\overline{\lambda}_{9}/\overline{\lambda})  +2(\overline{\lambda}_{10}/\overline{\lambda})= 2\,\overline{d}^{(2)}$ & $\overline{d}^{(3)} = (\overline{\lambda}_{14}/\overline{\lambda})/(16 \pi)\qquad\qquad\qquad\qquad$
        \\ \hline
$\overline{d}^{(4)} = [ 6 (\overline{\lambda}_{12}/\overline{\lambda}) + (5/4) (\overline{\lambda}_{13}/\overline{\lambda})]/\sqrt{4 \pi}\quad$ & $\overline{d}^{(5)} = [ 2 (\overline{\lambda}_{12}/\overline{\lambda}) + (1/4) (\overline{\lambda}_{13}/\overline{\lambda})]/\sqrt{4 \pi}\quad$ 
       \\ \hline
\end{tabular}
\label{TAB1}
\end{center}
\end{table}
Depending on the value of the slow-roll parameters the explicit 
evaluation of the various corrections follows in two complementary limits. The first limit is the one where $\epsilon$ is smaller than $1$ but not too small. Since the change of $\dot{\phi}$ during a Hubble time $H^{-1}$ 
follows from the background evolution\footnote{ In particular it follows from $2 (\overline{M}^2_{P}/M^2) \dot{H} = - \dot{\phi}^2$; this condition can obviously be rephrased as $ \dot{\phi}/H = \sqrt{2 \epsilon} \,\,\overline{M}_{P}/M$.} in this limit we can safely estimate that $M \simeq \sqrt{2 \epsilon} \,\overline{M}_{P}$. For generic theories of inflation (i.e. when $\varphi$ is {\em not} constrained by symmetry principles) $M$ cannot be much smaller than 
$\sqrt{2 \epsilon} \,\,\overline{M}_{P}$, otherwise $\dot{\phi}/H$ would diverge. 
If $M= \sqrt{2 \epsilon} \,\,\overline{M}_{P}$ then $H/M$ will be slightly larger than $H/\overline{M}_{P}$. 
In the case of conventional inflationary scenarios the physical wavenumber $k/a$ and the Hubble rate coincide at horizon exit and, more precisely, we will have that $\overline{M}_{P}^2 H^2/M_{P}^4 = \epsilon {\mathcal A}_{{\mathcal R}}/8$ where 
${\mathcal A}_{{\mathcal R}} = 2.41\times 10^{-9}$ is the amplitude of the curvature inhomogeneities assigned at the pivot scale $k_{p} = 0.002\, \mathrm{Mpc}^{-1}$ (see, for instance,\cite{THIRTEEN,RT0}).
If we keep track of the various factors the standard result $H/M_{P} = \sqrt{\pi \, \epsilon\, {\mathcal A}_{{\mathcal R}}}$ is readily obtained. Introducing then ${\mathcal A}_{0} = 8 \pi^3 {\mathcal A}_{{\mathcal R}} \simeq 6 \times 10^{-7}$ 
the leading contributions to $\chi_{E}^2$, $\chi_{B}^2$ and $\overline{\chi}^2$ are all ${\mathcal O}({\mathcal A_{0}}/\epsilon)$ 
while the subleading terms are\footnote{We are here considering that since $\eta = \epsilon - \overline{\eta}$ (with 
$\overline{\eta} = \overline{M}_{P}^2 (V_{,\varphi\varphi}/V)/2$), and since the scalar spectral index is 
$n_{s} = 1 - 6 \epsilon + 2 \overline{\eta}$, $\epsilon$ and $\eta$ can be approximately of the same order of magnitude.} 
 ${\mathcal O}(\eta\, {\mathcal A}_{0}/\epsilon)$ and ${\mathcal O}({\mathcal A}_{0})$.

The current observational determinations of the tensor to scalar ratio $r_{T}$ range between $r_{T} < 0.07$ \cite{RT0} and $r_{T}< 0.01$ \cite{RT1,RT2}. Since the consistency relations 
stipulate that $\epsilon \simeq r_{T}/16$, we have to acknowledge that  $\epsilon< 10^{-3} $ so that we are not in the situation discussed in the previous paragraph. For consistency we should then require, in the present context,  that 
$M \gg \sqrt{2\epsilon} \, \overline{M}_{P}$ which implies that $ M\simeq \overline{M}_{P}$ and $\epsilon \ll 1$. Consequently, the leading contributions appearing in Eqs. (\ref{ELEVEN}), (\ref{TWELVE}) and (\ref{THIRTEEN}) will be associated with $d_{E}^{(1)}$, $d_{B}^{(1)}$ and $\overline{d}^{(1)}$. The latter terms are all ${\mathcal O}(8 \pi {\mathcal A}_{0} \epsilon)$ while the former contain further powers of the slow-roll parameters. The suppressions coming from the inflationary evolution have to be combined with 
possible hierarchies of the different $d_{E}^{(i)}(\phi)$, $d_{B}^{(i)}(\phi)$ and $\overline{d}^{(i)}(\phi)$. All in all we two complementary situations emerge. In the first case the naturalness of the couplings would imply that all the $\lambda_{i}(\phi)$ are all of the order of $\lambda(\phi)$ and similarly 
for the $\overline{\lambda}_{i}(\phi)$ which should all be ${\mathcal O}(\overline{\lambda})$. In this situation the leading contribution to the gauge power spectra will be arguably given by the leading-order action. The same conclusion follows if $\lambda_{i}(\phi) \ll \lambda(\phi)$ and $\overline{\lambda}_{i}(\phi) \ll \overline{\lambda}(\phi)$. In the opposite situation $\lambda_{i}(\phi) \gg \lambda(\phi)$ and $\overline{\lambda}_{i}(\phi) \gg \overline{\lambda}(\phi)$ the hyperelectric and the hypermagnetic susceptibilities may evolve at different rates.  

Let us now make few concrete examples by first assuming the case of generic inflationary models and by positing, for the sake of simplicity, that all the $\lambda_{i}$ and $\overline{\lambda}_{i}$ are 
proportional to $\lambda$ and $\overline{\lambda}$ through some numerical 
constants of order $1$. In this case the coefficients of Tab. \ref{TAB1} become, in practice, $\phi$-independent and the 
leading-order expressions of the susceptibilities, as established above, is obtained 
by setting $M \sim \overline{M}_{P}$ and $\epsilon\ll 1$:
\begin{equation}
\chi_{X} = \sqrt{\frac{\lambda}{4\pi} }\,\,\sqrt{ 1 + \alpha_{X} \biggl(\frac{H}{M_{P}}\biggr)^2}, \qquad 
\overline{\chi}= \sqrt{\frac{\overline{\lambda}}{4\pi} }\,\,\sqrt{ 1 + \overline{\alpha} \biggl(\frac{H}{M_{P}}\biggr)^2}, 
\label{FOURTEEN}
\end{equation}
where $X = E,\,\, B$ so that $\alpha_{X}$ and $\overline{\alpha}$ do not depend on $\phi$.  As a second illustrative example we shall consider the more general version of the vertex considered in Ref. \cite{FOUR}
where a term $f(\phi) R_{\mu\alpha\nu\beta} \, Y^{\mu\alpha} \widetilde{\,Y\,}^{\nu\beta}$ 
has been considered in the context of polarized backgrounds of relic gravitons. This 
term corresponds to $\overline{\lambda}_{10}$  in Eq. (\ref{FIVE}) however, as we 
saw above, it does not make much sense to consider 
only $\overline{\lambda}_{10}$ for magnetogenesis considerations since $\overline{\lambda}_{8}$ a
and $\overline{\lambda}_{9}$ give exactly the same kind of contribution. For this reason 
as a further less generic model we could consider the case  
\begin{equation}
\chi_{E} = \chi_{B} = \sqrt{\frac{\lambda}{4\pi} }, \qquad \overline{\chi}= \sqrt{\frac{\lambda}{4\pi} }\sqrt{1 + q\,(2 - \epsilon) \biggl(\frac{H}{M_{P}}\biggr)^2},
\label{FOURTEENa}
\end{equation}
where $ q= {\mathcal O}(1)$ is just a numerical constant since we assumed, for the 
sake of simplicity, that all the $\lambda_{i}$ vanish while $\overline{\lambda}_{8} = \overline{\lambda}_{9} = \overline{\lambda}_{10} = \lambda = \overline{\lambda}$.
The various $\lambda_{i}$ and $\overline{\lambda}_{i}$ 
might also be much larger than $\lambda$ and $\overline{\lambda}$ and perhaps dominate the expressions 
of the susceptibilities. From the viewpoint of the underlying inflationary model it could also happen that the inflaton has some particular symmetry (like a shift symmetry $\varphi \to \varphi + c$) or that the rate of inflaton roll defined by $\eta$ remains
constant (and possibly larger than $1$), as it happens in certain fast-roll scenarios \cite{NON1} (see also, for instance, \cite{NON2,NON3}). In all these 
cases $\chi_{E}$ and $\chi_{B}$ may have rather different evolution and can be 
generically parametrized, in conformal time, as 
\begin{equation}
\chi_{E} = \biggl(- \frac{\tau}{\tau_{1}}\biggr)^{\gamma_{E}}, \qquad \chi_{B} = \biggl(- \frac{\tau}{\tau_{1}}\biggr)^{\gamma_{B}}.
\label{FIFTEEN}
\end{equation}
The non-generic classes of scenarios suggested by the present considerations 
can be multiplied and so far they not have been specifically analyzed. 

While examples of Eqs. (\ref{FOURTEEN}), (\ref{FOURTEENa}) and (\ref{FIFTEEN}) are only illustrative what matters, for 
the present ends, is that the general problem can be treated by adopting a new time parametrization and a consequent redefinition of the susceptibilities, namely:
\begin{equation}
\tau\to s = s(\tau), \qquad  d\tau= n(s)\, ds,\qquad n^2 = \chi_{E}^2/\chi_{B}^2, \qquad \chi = \sqrt{\chi_{E}\, \chi_{B}}.
\label{SEVENTEEN}
\end{equation}
It is relatively straightforward to rearrange Eqs. (\ref{TENa})--(\ref{TENb}) in the $s$-parametrization 
of Eq. (\ref{SEVENTEEN}) but probably the simplest way to discuss the problem without unnecessary details is to appreciate that in terms of $\chi$ and $n$ the comoving  fields are given by $\vec{B} = \vec{\nabla} \times \vec{{\mathcal A}}/\sqrt{n}$ 
and by $\vec{E} = - \, (\chi/\sqrt{n})\,\partial_{s} (\vec{{\mathcal A}}/\chi)$ where $\vec{{\mathcal A}}$ is the comoving vector potential defined in the Coulomb gauge \cite{VP} which is invariant under conformal rescaling.  If the latter expressions are inserted into Eq. (\ref{SIX}) the full action takes the following simple form:
\begin{equation}
S_{gauge}= \frac{1}{2} \int d^3 x\, \int \, ds \biggl[ \dot{{\mathcal A}}_{a}^2 + \biggl(\frac{\dot{\chi}}{\chi}\biggr)^2 {\mathcal A}_{a}^2  - 2  \biggl(\frac{\dot{\chi}}{\chi}\biggr) 
{\mathcal A}_{a} \, \dot{\mathcal A}_{a} - \partial_{i} {\mathcal A}_{a} \partial^{i} {\mathcal A}_{a} - {\mathcal C}(s) {\mathcal A}_{a} \partial_{b} {\mathcal A}_{m} \, \epsilon^{a \, b\, m} \biggr],
\label{EOM13}
\end{equation}
where the overdots now denote a derivation\footnote{The derivatives with respect to $s$ and the derivations with respect to the cosmic time coordinate $t$ never appear in the same context and, for this reason, we kept the overdot in the definitions of the slow-roll parameters (e.g. $ \epsilon= - \dot{H}/H^2$).}  with respect to the new time coordinate $s$. The canonical Hamiltonian associated with 
Eq. (\ref{EOM13}) easily follows. The classical fields and the conjugate momenta can then be promoted to the status of quantum operators so that the mode expansion of the hyperelectric and hypermagnetic fields in the circular basis turns out to be:
\begin{eqnarray}
&&\widehat{E}_{i}(\vec{x},s) = -  \int\frac{d^{3} k}{(2\pi)^{3/2} \, \sqrt{n(s)}} \sum_{\alpha=+,\,-} \, 
\biggl[ g_{k,\,\alpha}(s) \, \widehat{a}_{k,\alpha} \,\,  \varepsilon^{(\alpha)}_{i}(\hat{k})\,\,e^{- i \vec{k} \cdot\vec{x}} + g_{k,\,\alpha}^{\ast}(s) \, \widehat{a}_{k,\alpha}^{\dagger} \,\,  \varepsilon^{(\alpha)\ast}_{i}(\hat{k})\,\,e^{i \vec{k} \cdot\vec{x}}\biggr],
\label{EIGHTEEN}\\
&&\widehat{B}_{k}(\vec{x}, s) =  - i   \int\, \frac{\,\epsilon_{i\,j\,k}\, k_{i} \,d^{3} k}{(2\pi)^{3/2}\, \sqrt{n(s)}} \sum_{\alpha=+,\,-}\,
\biggl[ f_{k,\, \alpha}(s) \, \widehat{a}_{k,\,\alpha}\, \,\,  \varepsilon^{(\alpha)}_{j}(\hat{k})\, e^{- i \vec{k} \cdot\vec{x}} - f_{k,\, \alpha}^{\ast}(s) \, \widehat{a}_{k,\,\alpha}^{\dagger}\, \,\,  \varepsilon^{(\alpha)\ast}_{j}(\hat{k})\, e^{i \vec{k} \cdot\vec{x}}\biggr],
\label{NINETEEN}
\end{eqnarray}
where $\varepsilon^{(\pm)}(\hat{k})$  denote the two complex polarization obeying $\hat{k} \times \hat{\varepsilon}^{(\pm)} = \mp i\, \hat{\varepsilon}^{(\pm)}$; the creation and annihilation operators are directly defined in the circular basis and they obey the standard commutation relation $[\widehat{a}_{\vec{k}, \, \alpha}, \, \widehat{a}_{\vec{p}, \, \beta}] = \delta^{(3)}(\vec{k}- \vec{p})\, \delta_{\alpha\beta}$. The mode 
functions\footnote{It is easy to show that Eqs. (\ref{TWENTY}) also 
imply that $\ddot{f}_{k,\,\pm}+ [k^2 - \ddot{\chi}/\chi] f_{k,\,\pm} \mp C(s) k f_{k,\, \pm} =0$ 
while the first of the two equations becomes a definition of $g_{k,\,\pm}$, i.e.
 $  g_{k,\,\pm} = \dot{f}_{k,\, \pm} - (\dot{\chi}/\chi) f_{k,\, \pm}$.} appearing in Eqs. (\ref{EIGHTEEN}) and (\ref{NINETEEN}) obey:
\begin{equation}
\dot{f}_{k,\, \pm} = g_{k,\,\pm} + \biggl(\frac{\dot{\chi}}{\chi}\biggr) f_{k,\, \pm},\qquad
\dot{g}_{k,\,\pm} = - k^2 \, f_{k,\, \pm} - \biggl(\frac{\dot{\chi}}{\chi}\biggr) \,g_{k,\,\pm}  \pm  \,  {\mathcal C}(s)\, k \, f_{k,\, \pm}.
\label{TWENTY}
\end{equation}
From Eqs. (\ref{EIGHTEEN}) and (\ref{NINETEEN}) the two-point functions 
in Fourier space become:
\begin{eqnarray}
&& \langle \widehat{E}_{i}(\vec{k},s)\, \widehat{E}_{j}(\vec{p},s) \rangle = \frac{ 2 \pi^2 }{k^3} \biggl[\, P_{E}(k,s) \, p_{ij}(\hat{k}) 
+ P_{E}^{(G)}(k,s)\, \, i\, \epsilon_{i\, j\, \ell} \, \hat{k}^{\ell}\biggr] \, \delta^{(3)}(\vec{p} + \vec{k}),
\label{TWENTY1}\\
&& \langle \widehat{B}_{i}(\vec{k},s)\, \widehat{B}_{j}(\vec{p},s) \rangle = \frac{ 2 \pi^2 }{k^3} \biggl[\, P_{B}(k,s) \, p_{ij}(\hat{k}) 
+ P_{B}^{(G)}(k,s)\, \,i\, \epsilon_{i\, j\, \ell} \, \hat{k}^{\ell}\biggr] \, \delta^{(3)}(\vec{p} + \vec{k}),
\label{TWENTY2}
\end{eqnarray}
where $p_{i j} = \delta_{i j} - \hat{k}_{i} \, \hat{k}_{j}$ is the usual divergenceless projector.
In Eqs. (\ref{TWENTY1})--(\ref{TWENTY2}) $P_{E}(k,s)$ and $P_{B}(k,s)$ denote the hyperelectric and the hypermagnetic power spectra
while $P_{E}^{(G)}(k,s)$ and $P_{B}^{(G)}(k,s)$ are the corresponding gyrotropic contributions:
\begin{eqnarray}
P_{E}(k,s) &=& \frac{k^{3}}{4 \pi^2\, n(s)} \biggl[ \bigl| g_{k,\,-}\bigr|^2 + \bigl| g_{k,\,+}\bigr|^2 \biggr], \qquad
P_{B}(k,s) = \frac{k^{5}}{4 \pi^2\, n} \biggl[ \bigl| f_{k,\,-}\bigr|^2 + \bigl| f_{k,\,+}\bigr|^2 \biggr],
\label{TWENTY3}\\
P_{E}^{(G)}(k,s) &=& \frac{k^{3}}{4 \pi^2\, n(s)} \biggl[  \bigl| g_{k,\,-}\bigr|^2 - \bigl| g_{k,\,+}\bigr|^2 \biggr],\qquad
P_{B}^{(G)}(k,s) =  \frac{k^{5}}{4 \pi^2\, n(s)} \biggl[ \bigl| f_{k,\,-}\bigr|^2 - \bigl| f_{k,\,+}\bigr|^2\biggr].
\label{TWENTY4}
\end{eqnarray}
The results of Eqs. (\ref{SEVENTEEN})--(\ref{EOM13}) lead directly to Eqs. (\ref{TWENTY3})--(\ref{TWENTY4}) and 
are very convenient for estimating the 
magnitude of the corrections induced on the power spectra. To further illustrate the considerations developed so far we shall 
therefore analyze more specifically the cases of Eqs. (\ref{FOURTEEN}), (\ref{FOURTEENa}) and (\ref{FIFTEEN}).

We are now going to discuss some explicit solutions in those examples that we regard as more generic from the viewpoint 
of the theory.
Inserting Eq. (\ref{FOURTEEN}) into Eq. (\ref{SEVENTEEN}) we have that the effect of the generic corrections on $n$ and $\chi$ is ${\mathcal O}(H^2/M_{P}^2)$:
\begin{equation}
n = 1 + \frac{\alpha_{E} - \alpha_{B}}{2} \biggl(\frac{H}{M_{P}}\biggr)^2, \qquad \chi = \sqrt{\frac{\lambda}{4\pi}}\biggl[1 + \frac{\alpha_{E} + \alpha_{B}}{2} \biggl(\frac{H}{M_{P}}\biggr)^2\biggr].
\label{TWENTY5}
\end{equation}
We remind that we are here considering the situation where $\epsilon \ll 1$; this means that the corrections in Eq. (\ref{TWENTY5}) will be typically smaller 
than ${\mathcal O}(10^{-10})$. To deduce the correction on the power spectrum it is sufficient to solve Eq. (\ref{TWENTY}) directly in the $s$-parametrization; setting for simplicity $C(s)=0$ in the action (\ref{EOM13})  the WKB solution of Eq. (\ref{TWENTY}) is given by:
\begin{eqnarray}
f_{k}(s) &=& \frac{\chi(s)}{\chi_{ex}} \biggl\{ f_{k}(s_{ex}) 
+ \biggl[ \dot{f}_{k}(s_{ex}) - {\mathcal F}_{ex} f_{k}(s_{ex})\biggr] \int_{s_{ex}}^{s} \frac{\chi_{ex}^2}{\chi^2(s_{1})} \,d s_{1}\biggr\},
\nonumber\\
g_{k}(s) &=& \frac{\chi_{ex}}{\chi(s)} \biggl\{ g_{k}(\tau_{ex}) 
+ \biggl[ \dot{g}_{k}(\tau_{ex}) + {\mathcal F}_{ex} g_{k}(s_{ex})\biggr] \int_{s_{ex}}^{s} \frac{\chi^2(s_{1})}{\chi_{ex}^2} \,d s_{1}\biggr\},
\label{TWENTY6}
\end{eqnarray}
where ${\mathcal F}_{ex} = (\dot{\chi}/\chi)_{ex}$. Since the leading terms in Eq. (\ref{TWENTY6}) are given by 
$(\chi(s)/\chi_{ex} )\,f_{k}(s_{ex})$ and by $(\chi_{ex}/\chi(s)) \, g_{k}(s_{ex})$,  for typical wavelengths 
larger than the Hubble radius during inflation Eq. (\ref{TWENTY5}) implies that $P_{B}(k, s)= \overline{P}_{B}(k,\tau) [1 + {\mathcal O}(H^2/M_{P}^2)]$ where $\overline{P}_{B}(k,\tau)$ is the spectrum obtained when $\alpha_{B} = \alpha_{E} =0$. 
A conclusion similar to the one of Eqs. (\ref{TWENTY5})--(\ref{TWENTY6}) follows after inserting Eq. (\ref{FOURTEENa}) into Eq. (\ref{TWENTY}):
\begin{equation}
\ddot{f}_{k,\, \pm} + \biggl\{ k^2 - \frac{\ddot{\sqrt{\lambda}\,\,\,\,}}{\sqrt{\lambda}} \mp k\, \biggl[ \frac{\dot{\lambda}}{\lambda}  - 2 q \epsilon \biggl(\frac{H}{M_{P}}\biggr)^2 \, {\mathcal H}\biggr]\biggr\} \, f_{k,\,\pm}=0.
\label{TWENTY6a}
\end{equation}
This equation is nothing but the standard equation for the Whittaker's functions \cite{abr}. Indeed by rescaling the 
coordinates as $z = 2 \, i \, k$,  Eq. (\ref{TWENTY6a}) becomes\footnote{Note, as a side remark, 
that in this particular example $s$ and $\tau$ coincide exactly.}:
\begin{equation}
\frac{d^2 f_{k, \, \pm}}{ dz^2} + \biggl\{ - \frac{1}{4} - \frac{\mu^2 -1/4}{z^2} + \frac{\zeta}{z}\biggl[ 1 + 2 q \frac{\epsilon^2 {\mathcal A}_{R}}{2 \mu+ 1}\biggr] \biggr\}  f_{k, \, \pm}=0,
\label{TWENTY6b}
\end{equation}
where $\zeta = i\, (\mu+1/2)$. In Eq. (\ref{TWENTY6b}) we assumed a power-law dependence for $\sqrt{\lambda}$ but the relevant 
aspect concerns the comparison of the two terms in the squared bracket. Since $\epsilon \ll 1$ (and typically ${\mathcal O}(10^{-3})$)
the second term is completely negligible with respect to $1$: a simple estimate implies that $\epsilon^2 {\mathcal A}_{{\mathcal R}} < 10^{-15}$. Note, in this respect, that the effect goes as $\epsilon^2$ since the first $\epsilon$ comes from the derivative 
of $H^2$ while the second one follows from the numerical value of $H$ in terms of ${\mathcal A}_{{\mathcal R}}$. 

Let us finally come to Eq. (\ref{FIFTEEN}) which is interesting since it can be realized in the context of some non-generic models of inflation and anyway when the $\lambda_{i}(\phi) \gg {\mathcal O}(\lambda)$.
As before, inserting Eq. (\ref{FIFTEEN}) into Eq. (\ref{SEVENTEEN}) we obtain, after simple algebra,
\begin{equation}
\biggl(- \frac{\tau}{\tau_{1}}\biggr)^{\gamma_{B} - \gamma_{E} +1} = \biggl(- \frac{s}{s_{1}}\biggr), \qquad n(s) = \biggl( - \frac{s}{s_{1}}\biggr)^{\frac{\gamma_{E} -\gamma_{B}}{(\gamma_{B}- \gamma_{E} +1)}}, \qquad \chi(s) = \chi_{1} \biggl( - \frac{s}{s_{1}}\biggr)^{\frac{\gamma_{E} +\gamma_{B}}{2(\gamma_{B}- \gamma_{E} +1)}},
\label{TWENTY7}
\end{equation}
where $s_{1} = \tau_{1} /(1 - \gamma_{E} + \gamma_{B})$. In what follows we shall assume $\gamma_{B}>0$ and $\gamma_{E}>0$. 
The solution of the evolution for the mode functions during the inflationary stage follows 
from Eq. (\ref{TWENTY}) and it can be directly obtained in the $s$-parametrization: 
\begin{equation}
f_{k}(s) = \frac{N_{\mu}}{\sqrt{ 2 k}} \sqrt{ - k \, s}\, H_{\mu}^{(1)}(- k\,s), \qquad g_{k}(s) = - N_{\nu} \sqrt{\frac{k}{2}} \sqrt{- k\,s} H_{\nu}^{(1)}(- k\,s)
\label{TWENTY8}
\end{equation}
where $| N_{\mu} | = |N_{\nu}| = \sqrt{\pi/2}$ while $\mu= | (2 \gamma_{E}-1)/[ 2 (\gamma_{B}- \gamma_{E} +1)]\,|$ and 
$\nu= ( 2 \gamma_{B} + 1)/[ 2 (\gamma_{B}- \gamma_{E} +1)]$ (note the absolute value in the expression of $\mu$). From Eqs. (\ref{TWENTY1}) 
and (\ref{TWENTY8}) the inflationary power spectra easily follow and they are
\begin{equation}
P_{B}(k,s,\tau) =\frac{a^4 H^4}{n(s)} \biggl(\frac{\tau}{s}\biggr)^4 D(\mu) (- k\, s)^{5 - 2 \mu}, \qquad 
P_{E}(k,s,\tau) =\frac{a^4 H^4}{n(s)} \biggl(\frac{\tau}{s}\biggr)^4 D(\nu) (- k\, s)^{5 - 2 \nu}, 
\label{TWENTY9}
\end{equation}
where, for a generic argument $z$, $D(z) = 2^{2 z-3}\,\Gamma^2(z)/\pi^3$. Note that the two power spectra can be 
usefully viewed, for practical purposes, as functions of $\tau$ and $s$; depending on the specific necessity Eq. (\ref{TWENTY7})
will be used to eliminate one of the two time variables. Various considerations restrain the variability of $\gamma_{E}$ and $\gamma_{B}$;
for instance to have $\tau^4/[ n(s) \, \, s^4] < 1$ throughout the whole inflationary stage we must require $\gamma_{E} < \gamma_{B}$;
to have $\chi$ increasing during inflation we mis demand $\gamma_{E} < \gamma_{B}+1$; finally to 
avoid that the electric and magnetic fields will be overcritical during inflation we must have $\gamma_{E} < (3\gamma_{B}/+ 4)/5$.
\begin{figure}[!ht]
\centering
\includegraphics[height=7cm]{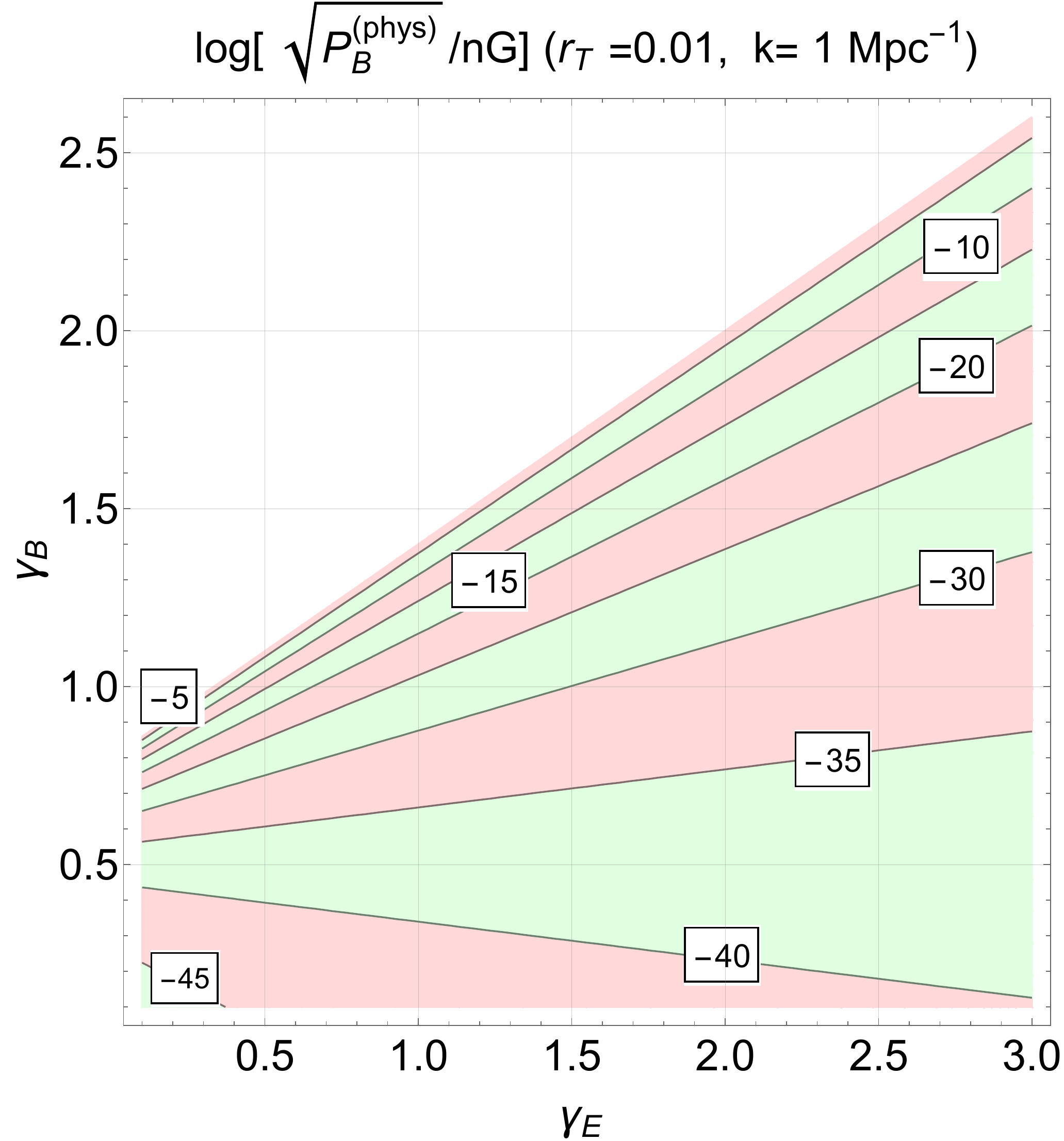}
\includegraphics[height=7cm]{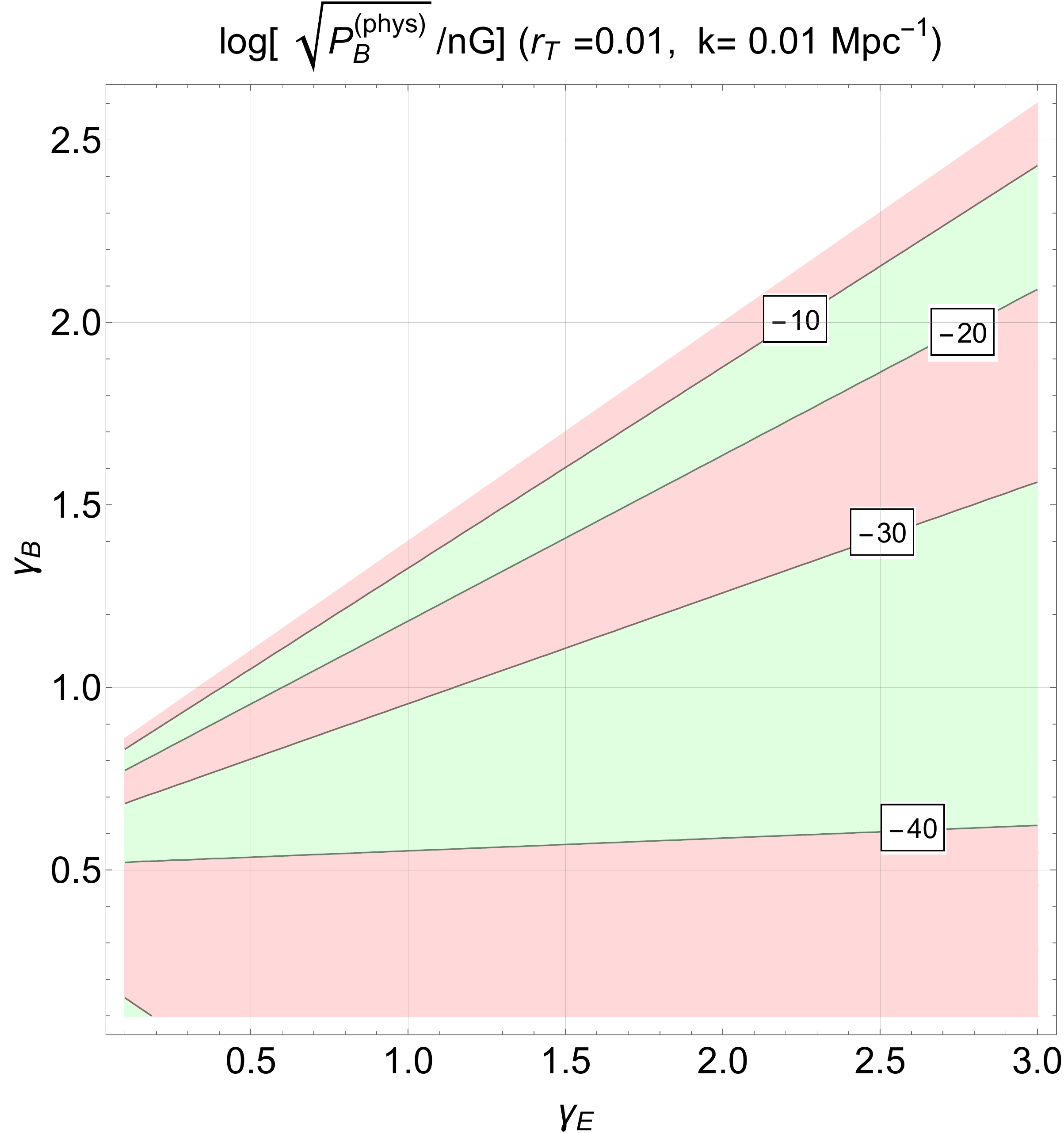}
\caption[a]{The common logarithm of the 
power spectrum is illustrated in the plane $(\gamma_{E}, \,\gamma_{B})$ for $r_{T}=0.01$ and for two different 
scales i.e. $k= 1\, \mathrm{Mpc}^{-1}$ (plot at the left) and and $k =0.01\, \mathrm{Mpc}^{-1}$ (plot at the right). }
\label{FIG1}      
\end{figure}
Taking into account all the relevant constraints the physical magnetic fields after inflation can be explicitly evaluated 
when the relevant scales reenter the Hubble radius, i.e. for $\tau_{k} \simeq 1/k$ where $k$ is of the order 
of the $\mathrm{Mpc}^{-1}$ which is the typical scale for magnetogenesis considerations:
\begin{equation}
P_{B}^{(phys)}(k, \tau_{k}) \simeq H_{1}^4 \biggl(\frac{a_{1}}{a_{k}}\biggr)^4 {\mathcal M}(\gamma_{E},\gamma_{B}) \biggl(\frac{k}{a_{1} H_{1} }\biggr)^{\alpha(\gamma_{E},\gamma_{B})}, \qquad \alpha(\gamma_{E},\gamma_{B}) = ( 4 - 5 \gamma_{E} + 3\gamma_{B})/(1 -\gamma_{E} + \gamma_{B})
\label{THIRTY}
\end{equation}
where ${\mathcal M}(\gamma_{E},\gamma_{B})$ is a numerical factor that varies between $0.2$ and $5 \times 10^{-3}$ 
when $0<\gamma_{B}<4$ and $\gamma_{E}< (3\gamma_{B} +4)/5$. In Fig. \ref{FIG1} the upper right corner 
is in fact excluded by the critical density constraint; in the remaining parts of the plots we illustrated 
the common logarithm of $\sqrt{P_{B}^{(phys)}(k, \tau_{k})}$ expressed in nG. 
\begin{figure}[!ht]
\centering
\includegraphics[height=7cm]{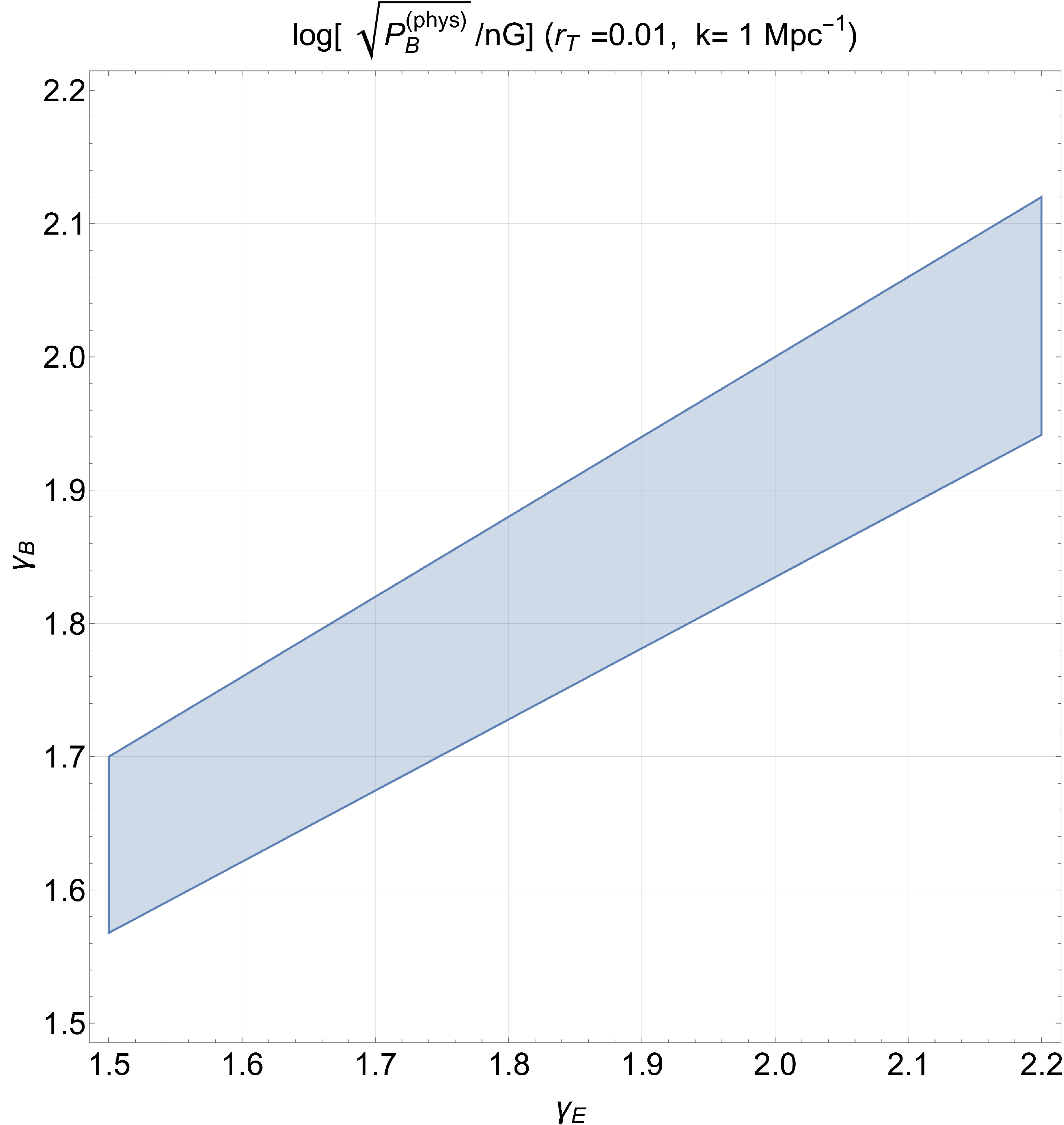}
\includegraphics[height=7cm]{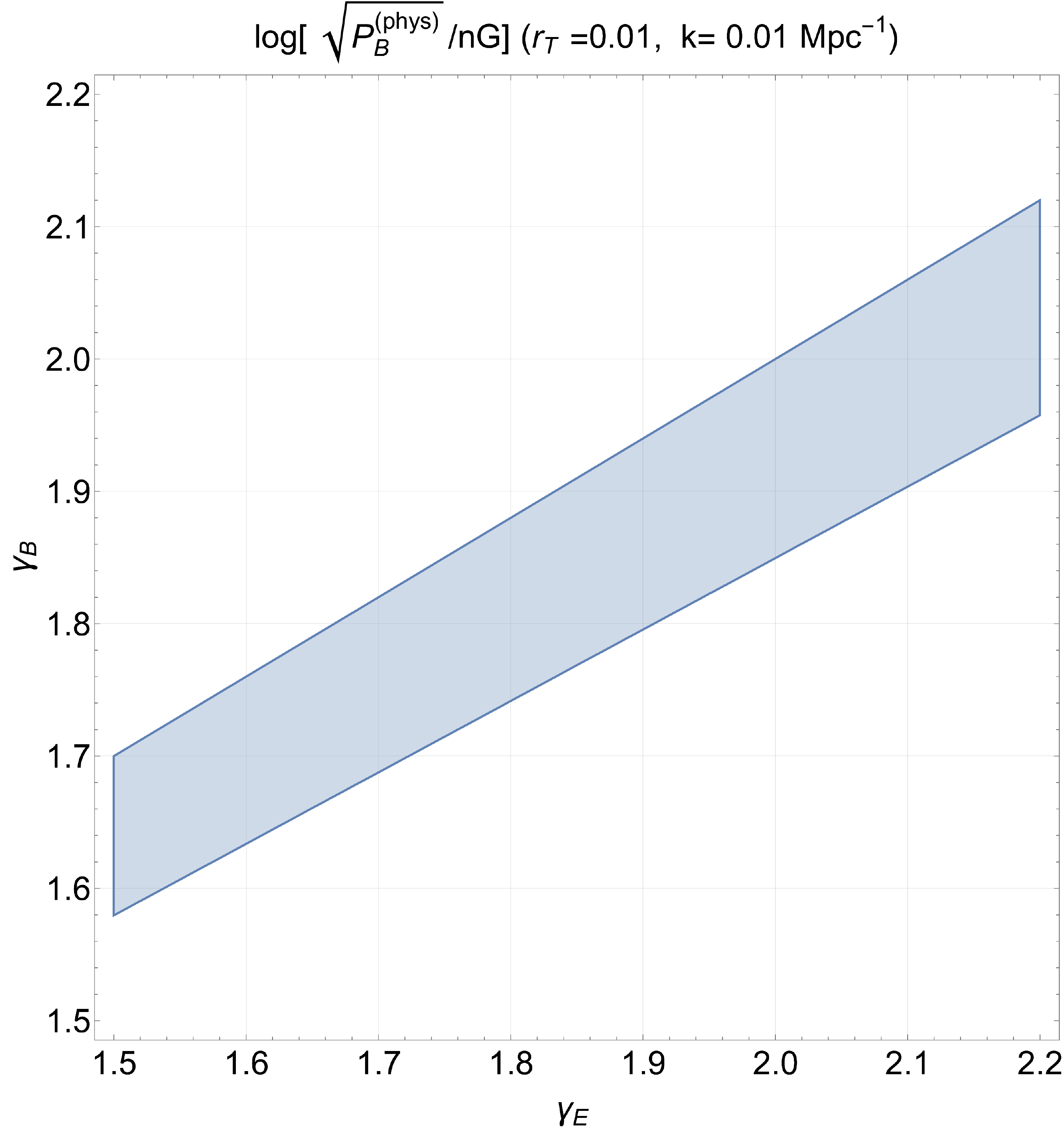}
\caption[a]{The allowed region of the parameter space is illustrated with a 
shaded area for fixed $k$ and in the $(\gamma_{E}, \, \gamma_{B})$ plane. The shaded regions correspond to the 
are where all the phenomenological requirements are satisfied.}
\label{FIG2}      
\end{figure}
To achieve a successful magnetogenesis the least demanding requirement 
(i.e. $\sqrt{P^{(phys)}_{B}} \geq 10^{-16}\,\mathrm{nG}$) follows by assuming 
that, after compressional amplification, every rotation of the galaxy increases the initial magnetic field of one $e$-fold. According to some this requirement is not completely reasonable since it takes more than one $e$-fold  to increase the value of the magnetic field by one order of magnitude  and this is the rationale for the most demanding condition i.e. $\sqrt{P^{(phys)}_{B}} \geq 10^{-11}\,\mathrm{nG}$.
In Fig. \ref{FIG2} the shaded areas denote the region where the spectral energy density 
is subcritical both during and after inflation while the magnetogenesis 
and the Cosmic Microwave Background constraints 
are all satisfied. In a conservative perspective we
required that the physical power spectrum after equality (but before decoupling )
is smaller than $10^{-2}$ nG for typical wavenumbers comparable with the pivot 
scale $k_{p} = 0.002\, \mathrm{Mpc}^{-1}$ at which the scalar and tensor power spectra 
are customarily assigned (see e.g. \cite{FOURa}).  In the two plots of Fig. \ref{FIG2}, always in a conservative perspective, we required $\sqrt{P^{(phys)}_{B}} \geq 10^{-11}\,\mathrm{nG}$. It is finally interesting to remark that the allowed region of Fig. \ref{FIG2} naturally selects models 
that satisfy the approximate condition $\gamma_{E} < \gamma_{B}$ implying 
that $n > 1$. 

Let us conclude with some  comments and caveats on the overall logic of the present analysis. Since a generic spectator field $\psi$ may replace the inflaton and lead to plausible magnetogenesis scenarios (see \cite{FOURa} and references therein),
the considerations developed here also apply when the various $\lambda_{i}$ and $\overline{\lambda}_{i}$ are $\psi$-dependent quantities. The contributions with four derivatives remain essentially the same but must be considered in conjunction with the supplementary restrictions associated with the different physical nature of the spectator fields. Furthermore if the couplings depend simultaneously on the inflaton $\phi$ and on $\psi$ further terms (containing the gradients of $\psi$) will have to be included in the effective Lagrangian. The results of these additions will not crucially modify the general structure of the gauge action that will be always diagonalized in the $s$-time parametrization. There will certainly  be  some physical differences since the spectator fields induce entropic fluctuations that may also affect the consistency relations as well as other aspects of the CMB initial conditions \cite{FOURa}: both themes are beyond the present discussion but have been analyzed in the past with fairly general conclusions (see e. g. \cite{FOURb}). We also remark that for $\epsilon \ll 1$ and $M\simeq \overline{M}_{P}$ the effect of the parity violating contributions coming from the corrections is always subleading  and this result generalises the earlier discussions of Ref. \cite{SIX} where the coupling to the inflaton and the parity-violating terms of the type \cite{FOUR} have been neglected altogether. If the higher-order corrections are non-generic either because of some symmetry of the inflaton or because of specific dynamical assumptions (like in the case of fast-roll models \cite{NON1,NON2,NON3}), it is plausible that the hyperelectric and the hypermagnetic susceptibilities will evolve at a different rate. All in all the systematic approach discussed here can be productively applied to slightly different situations while the overall logic and the general results will remain unchanged. 

It is a pleasure to thank T. Basaglia and S. Rohr of the CERN Scientific Information Service for their 
kind and reliable assistance.

\end{document}